\def\ka{\hbox{\ae}}
\documentclass[a4paper]{article}
\usepackage{amssymb,amsmath}
\usepackage{graphics,epsf}
\usepackage{floatflt}

\renewcommand{\Re}{\hbox{Re}\hspace{.2pt}}

\title{\bf Interference at  quantum transitions:
lasing without inversion and resonant four-wave mixing  in strong
fields at Doppler-broadened transitions}
\author{ A. K. Popov\\ 
Institute for Physics,
Russian Academy of Sciences, Krasnoyarsk University
and Krasnoyarsk\\ Technical University 660036 Krasnoyarsk, Russia, E-mail: popov@ksc.krasn.ru
}
\date{}
\begin{document}
\voffset=-30mm
\hoffset=-20mm
\maketitle
\sloppy
\setcounter{page}{252}
\begin{abstract}
An influence of nonlinear interference processes at quantum transitions under
strong resonance electromagnetic fields on absorption, amplification and
refractive indices as well as on four-wave mixing processes is investigated.
Doppler broadening of the coupled transitions, incoherent excitation,
relaxation processes, as well as  power saturation processes associated with
the coupled levels are taken into account. Both closed (ground state is
involved) and open (only excited states are involved) energy level
configurations are considered. Common expressions are obtained which allow one
to analyze the optical characteristics (including gain without inversion and
enhanced refractive index at vanishing absorption) for various $V$, $\Lambda$
and $H$ configurations of interfering transitions by a simple substitution of
parameters. Similar expressions for resonant four-wave mixing in Raman
configurations are derived too. Crucial role of Doppler broadening is shown.
The theory is applied to numerical analysis of some recent and potential experiments.

{\bf Keywords:} quantum interference, lasing without inversion,
resonant four-wave mixing,  Doppler and strong field  effects.
{PACS: 42.50.Gy, 42.55.-f, 42.65}
\end{abstract}

\section{\small INTRODUCTION}
Many concepts of quantum optics were originated proceeding from the assumed
equality of probabilities of induced transitions accompanied by an absorption
and emission of photons predicted by A. Einstein. Requirement of population
inversion for lasing is direct consequence from this equality.

At the presence of several resonant electromagnetic fields, probability
amplitudes of a coupled quantum states contain several oscillating components
at close frequencies. Therefore alongside with squared modules of appropriate
components cross terms indicating an interference of quantum transitions appear
while calculating transitions probabilities. The coherent nonlinear
optical phenomena stipulated by the indicated evolution of quantum
states, driven by several fields, were called as nonlinear interference
effects ($NIE$)~\cite{{Shalag},{Vved}}. In quantum optics $NIE$ may result in
different coupling of a radiation with atoms in absorbing and emitting states
controlled by the auxiliary fields \cite{{Not},{Sok}}. Various appearances of these
effects are feasible.  Soon after discovery of lasers Rautian and Sobelman~\cite{Sobel}
showed feasibility of amplification without inversion ($AWI$) in two-level systems.
The features of $AWI$ in optical three-level systems were explored in \cite {{Pop},{Vved}}.
Studies of $NIE$ in absorption/gain spectra including experiments on
generation of an optical radiation in three-level systems, so that generation
was possible  only at the expense of nonlinear interference effects, drew much
attention in 60th and 70th \cite {Bet}. (Review of relevant optical experiments
and of earlier papers on $AWI$ in microwave range see in \cite {{Spie},{Izv}}.)
Coherent population trapping ($CPT$) is one of the appearances of $NIE$ for the
states with negligible relaxation rates and Doppler effects. In 80th -- 90th
studies of coherent interference processes at
quantum transitions have been attracting much interest again in the context of
$AWI$, electromagnetically - induced transparency ($EIT$), $CPT$, enhanced
nonlinear optical frequency conversion  and other manifestations of these
effects \cite {Scul}.

In classical terms an emission and absorption of a radiation are stipulated by
forced oscillations of bound charges and depend on phase difference between
radiation and induced oscillations.  However,  a radiation may simultaneously
drive several coherent interfering oscillations of a various origin. Depending
on a relation of their  phases and amplitudes the interference can be either
constructive or destructive, full or partial.  Thus the matching components of
an optical response can either amplify or suppress each other.  On the other
hand, the macroscopic response of a medium can be thought as result of quantum
transitions, at which the photons can \\
\\
{\footnotesize
11th International Vavilov Conference on Nonlinear Optics, 24-28 June 1997,
Novosibirsk, Russia}\hfill {\bf SPIE Vol. 3485}

\newpage\noindent simultaneously contribute in several quantum pathways.
By applying semi-classical approach a deep analogy with many well known
effects of classical physics can be used to interpret and foresee the
relevant quantum optics effects. Thus, leaving aside classifications of
involved elementary quantum processes (introduced and valid for weak
fields in the limits of perturbation theory), it is possible to predict
and to explain  wide range of  optical processes, stipulated by quantum
interference, some of them are quite unusual.

The objective of the paper is to consider various appearance of interference effects
in resonance nonlinear - optical  processes  with the aid of the outlined approach in
a context of some recent experiments.  The amplitude and phase relations of
interfering intra-atomic oscillations depend on configuration and on relaxation
characteristics of the coupled transitions, on type of nonlinear-optical process, on
intensities and frequency detunings of the radiations from resonances. Due to the
difference in Doppler shifts the contributions to the macroscopic polarization of
atoms at various velocities in gases may interfere in a different way too. The
interference appears differently in an absorption, refraction and in different
four-wave mixing ($FWM$) processes.

As an illustration of $NIE$  the following results will be presented:

1.  The possibility of an amplification of a radiation without inversion of
saturated populations on resonant transition is investigated.  Influence  of
the growth of intensity of an amplified radiation on inversionless
amplification in various open and closed transition configurations
is analyzed.  The conditions are formulated and with the concrete
examples is shown, that by proper change of incoherent excitation rate
of levels and of auxiliary radiation intensity the index of an inversionless
amplification does not decrease with growth of intensity of an amplified
radiation. The elements of the theory of such lasing without inversion are
presented.

2. It is shown, that due to Maxwell velocity distribution of atoms and
corresponding inhomogeneous broadening of the coupled transitions, incoherent
excitation of the intermediate levels may drastically change both spectral
properties and a magnitude (by orders of magnitude) of nonlinear susceptibilities for
resonant $FWM$ processes. As the consequence, important power saturation
effects appear. These features must be taken into account  for explanation of
the experiments and optimization of frequency-conversion. Resonant $FWM$ coupling of two
strong and two weak radiations is considered. Formulas for both cases of
coupling, one is relevant to coherent population trapping, another --  when
each level is coupled to only one driving field are derived.  The outcomes are
applied to numerical analysis and to discussion of recent experiments \cite
{Hin}.

\section{\label{sec1}\small ABSORPTION AND REFRACTION INDICES
FOR A STRONG RADIATION AT THE PRESENCE OF OTHER STRONG RADIATION,
COUPLED TO AN ADJACENT TRANSITION}

\begin{floatingfigure}[l]{60mm}
\epsfxsize=50mm \center{\leavevmode\epsfbox{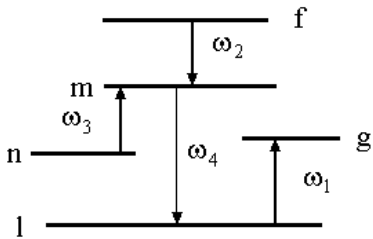}}
\vspace{-1mm}\caption{\label{sh}\small Transition configurations. }
\end{floatingfigure}

First,
consider interaction of two strong laser fields at the three-level system. Possible
configurations of such systems are shown in FIG.\ref{sh}: folded $ V $ and $ \Lambda
$, and  cascade - $ H $.  In further we shall investigate  spectral features of a gas
material for a radiation $ E _ 4 $ at frequency $ \omega _ 4 $, tunable in the
vicinity of a transition  $ l-m $. It's intensity is not supposed weak. Depending on
the configuration of transitions under consideration  one of the auxiliary strong
fields $ E_1 $, $ E _ 3 $ or $ E _ 2 $ with frequencies $ \omega _ 1 $, $ \omega _ 3
$, $ \omega _ 2 $, resonant to adjacent transitions shown in the figure is turned
on.  All radiations are supposed to be uniformly polarized co- or counter-propagating
travelling wave: $E_{j}(z,t)=E_{j}\exp \{-[i(\omega_{j}t-k_{j}z)]\}+ k.c., $
where $ k _ {j} $ - can take both positive and negative values,
$j=1,2,3,4$.  Incoherent excitation of the levels with Maxwell's velocity
distribution, all possible population  and coherence relaxation channels are
accounted for.

It is necessary to distinguish the open and closed energy-level configurations.
In open one (lowest level is not ground), the rate of incoherent excitation of
the levels by an external source practically does not depend on the rate of
induced transitions between considered levels.  In the closed one (lowest level
is ground one), the excitation rate for atoms at different levels and
velocities depends on the value and velocity distribution of the other
populations, which are dependent on the intensity of the driving fields.
\subsection {General equations for  absorption end refraction indices}
Power dependent susceptibility $\chi_{4}$, responsible for absorption and
refraction, can be found from the equation:
\begin{equation}
P^{NL}(\omega _{4})=N \chi_{4}E_{4},
\end{equation}
where polarization $P^{NL}(\omega _{4})$ is convenient to calculate with aid of
density matrix $\rho _{ij}$:
\begin{equation}
{P}=N\rho _{ij} d_{ji}+{ c. c.}.
\end{equation}
\noindent
Taking into account above discussed relaxation and incoherent excitation
processes, density matrix equations for a mixture of pure quantum mechanical
ensembles in the interaction representation can be written in general form as:
\begin{eqnarray}
L_{nn}\rho_{nn}=q_n-i[V,\rho]_{nn}+ \gamma_{mn}\rho_{mm},
L_{lm}\rho_{lm}=L_4\rho_4=-i[V,\rho]_{lm},\label{ro}\\
L_{ij}=d/dt\,+\Gamma_{ij},\quad
V_{lm}=G_{lm}\cdot\exp\{i[\Omega_4t - k z]\},\quad
G_{lm}=-{\bf E_4\cdot d_{lm}}/2\hbar,\nonumber
\end{eqnarray}
where $ \Omega _ 4 = \omega _ 4-\omega_ {mn} $ - frequency detuning from
resonance; $ \Gamma _ {mn} $ - homogeneous half-widths of transitions, in
absence of collisions $ \Gamma _ {mn} = (\Gamma_m + \Gamma_n)/2$; $ \Gamma _ n
 = \sum _ j\gamma _ {nj} $ - inverse lifetimes of levels; $ \gamma _ {mn} $ -
rate of relaxation from the level $m$ to $n$, $ q _ n = \sum _ j w _ {nj} r _ j
$ - rate of incoherent excitation to a state $ n $ from underlying levels.  For
open configurations $ q _ i $ - is mainly determined by the population of the
ground state and practically does not depend on the driving fields.

In a steady-state regime a set of density-matrix equation  may be reduced to
the set of algebraic equations \cite{Kuch}. Below we present only results of
calculations.  Despite of essential distinctions in manifestations of $NIE$ in
different open and closed configuration, formulas for  absorption/gain
($\alpha$) and resonant part of refractive ($\delta n$) indices  and also for
power dependent populations of the levels  can be presented uniformly for all
configurations shown on FIG.\ref{sh}:
\begin{eqnarray}\label{alpharch}
&\alpha_4/\alpha_{04}=Re\{\chi_4/\chi_4^0\},\>
\delta n_4/2\delta n_{04}=Im\{\chi_4/\chi_4^0\},\>
\alpha_i/\alpha_{0i}=Re\{\chi_i/\chi_i^0\},\>
\delta n_i/2\delta n_{0i}=Im\{\chi_i/\chi_i^0\},&\nonumber\\
&\dfrac{\chi_4}{\chi_4^0}=\dfrac{\Gamma_4}{P_4}\dfrac{\Delta r_4(1+u_2)
\mp \Delta r_i g_2 }{\Delta n_4(1+g_1+u_2)},\qquad
\dfrac{\chi_i}{\chi_i^0}=\dfrac{\Gamma_i}{P_i}\dfrac{\Delta r_i(1+g_1^*)
\mp \Delta r_4 u_1^* }{\Delta n_4(1+g_1^*+u_2^*)},&
\end{eqnarray}
Here and further index $ i $ specifies transition, resonant to the auxiliary
radiation (see FIG.\ref{sh}), $\chi$ is susceptibility, $\alpha_0, \delta n_0,
\chi_0$ are corresponding maximum resonant values at zero field intensities, $
P _ j = \Gamma _ j + i\Omega _ j $ (for example: $ P _ {lm} = P _ 4 = \Gamma _
4 + i\Omega _ 4 $, $ P _ {lm} = P _ {ml} ^ * $, $ P _ {lf} = P _ {42} = \Gamma
_ {lf} + i (\Omega _ 4 + \Omega _ 2) $ etc.).  If the atom moves with speed $ v $,
Doppler shift of resonances must be taken into account by substitution  $
\Omega _ j $ for $ \Omega _ j ^ {'} = \Omega _ j - k _ jv $.  In further
strokes will be omitted, but it is supposed, that the Doppler shift in the
formulas is taken into account.
$\Delta r _ 4 = r _ l-r _ m $ is power dependent population difference; $
\Delta n _ 4 = n _ l-n _ m $, $ n _ i $ - population of the level in absence of
driving  fields, which is described by the formula: $ n _ i = (q _ i/\Gamma _
i) + (\gamma _ {ki} / {\Gamma _ i}) ({q _ k} / {\Gamma _ k}) $.
\begin{eqnarray}\label{g}
g_1={|G_i|^2}/{P_4P_{4i}},
g_2={|G_i|^2}/{P_i^*P_{4i}},
u_1={|G_4|^2}/{P_4P_{4i}},
u_2={|G_4|^2}/{P_i^*P_{4i}}.
\end{eqnarray}
$G_j$ are coupling Rabi frequencies. Formulas for populations differences
can be presented uniformly too:
\begin{eqnarray}\label{der}
&\Delta r_4=(\Delta n_4X_{2}\mp\Delta n_iX_{3})/(X_{1}X_{2}-X_{3}X_{4}),
 \Delta r_i=(\Delta n_iX_{1}\mp\Delta n_4X_{4})/(X_{1}X_{2}-X_{3}X_{4}).&\\
&X_{2}=1+Re\{a_{24} \ka_4 \dfrac{\Gamma_4}{P_4}\dfrac{g_2 }{1+g_1+u_2}
+a_{2i} \ka_i \dfrac{\Gamma_i}{P_i}\dfrac{1+g_1^*}{1+g_1^*+u_2^*}\},&\nonumber\\
&X_{3}=Re\{a_{34} \ka_4 \dfrac{\Gamma_4}{P_4}\dfrac{g_2 }{1+g_1+u_2}
+a_{3i} \ka_i \dfrac{\Gamma_i}{P_i}\dfrac{1+g_1^*}{1+g_1^*+u_2^*}\},&\nonumber\\
&X_{1}=1+ Re\{a_{14} \ka_4\dfrac{\Gamma_4}{P_4}\dfrac{1+u_2}{1+g_1+u_2}
+a_{1i} \ka_i \dfrac{\Gamma_i}{P_i}\dfrac{u_1^* }{1+g_1^*+u_2^*}\},&\nonumber\\
&X_{4}=Re\{a_{44} \ka_4 \dfrac{\Gamma_4}{P_4}\dfrac{1+u_2}{1+g_1+u_2}
+a_{4i}\ka_i\dfrac{\Gamma_i}{P_i}\dfrac{ u_1^* }{1+g_1^*+u_2^*}\},&\nonumber\\
&a_{14}=-a_{34}, a_{1i}=-a_{3i}, a_{24}=-a_{44}, a_{2i}=-a_{4i}.&
\end{eqnarray}
Sign  minus in (\ref{alpharch}),(\ref{der}) concerns to folded $ V $ ($ E _ 4,
E _ 1 $) and $ \Lambda $ ($ E _ 4,E _ 3 $) schemes, plus - to cascade $ H $ ($
E _ 4, E _ 2 $) scheme. Beside that in the ladder $H$-scheme, $P_i$ must be
substituted for $P_i^*$, and vice a versa: $P_i^*$ for $P_i$.  $ \ka _ 4 $ and
$ \ka _ i $ - are saturation parameters accordingly for transitions $4 $ and $
i $. For open configurations
\begin{equation}\label{ka}
\ka_4={2|G_4|^2(\Gamma_l+\Gamma_m-\gamma_4)}/{(\Gamma_l\Gamma_m\Gamma_4)},
\end{equation}
whereas $ \ka _ i $  and  parameters $ a _ {ij} $ depending only on relaxation
constants are defined below for each configuration.\\

\centerline{\bf 1. $ V $ - scheme (fields $ E _ 4 $, $ E _ 1 $; $ i = 1 $)}
\noindent {OPEN CONFIGURATION}
\begin{eqnarray}\label{vo}
&\ka_i=\ka_1=\dfrac{2|G_1|^2(\Gamma_g+\Gamma_l-
\gamma_1)} {\Gamma_g\Gamma_l\Gamma_1},
a_{2i}=a_{21}=1,  a_{14}=1,
a_{3i}=a_{31}= \dfrac{\Gamma_g-\gamma_1}{\Gamma_g+\Gamma_l-\gamma_1},
 a_{44}=\dfrac{\Gamma_m-\gamma_4}{\Gamma_l+\Gamma_m-\gamma_4}.&
\end{eqnarray}
\noindent{CLOSED CONFIGURATION}
\begin{eqnarray}\label{vc}
&
\ka_4= {4|G_4|^2}/{\Gamma_m\Gamma_4},
\ka_i=\ka_1= {4|G_1|^2}/{\Gamma_g\Gamma_1},&\nonumber\\
&
a_{3i}=a_{31}=0.5\Delta n_4, a_{44}=0.5\Delta n_1,
a_{2i}=a_{21}=0.5[1+\Delta n_1],  a_{14}=0.5[1+\Delta n_4].
&
\end{eqnarray}
\centerline{\bf 2. $ \Lambda $ - scheme (fields $ E _ 4 $, $ E _ 3 $, $ i =
3 $)}
\noindent {OPEN CONFIGURATION}
\begin{eqnarray}\label{lo}
&\ka_3={2|G_3|^2(\Gamma_m+\Gamma_n-\gamma_3)}/
{\Gamma_m\Gamma_n\Gamma_3},\quad
a_{2i}=a_{23}=1,\quad a_{14}=1,&\nonumber \\
&
a_{3i}=a_{33}={\Gamma_n}{(\Gamma_l-\gamma_4)}/{\Gamma_l}
{(\Gamma_m+\Gamma_n-\gamma_3)},\quad
a_{44}={\Gamma_l}{(\Gamma_n-\gamma_3)}/{\Gamma_n}
{(\Gamma_m+\Gamma_l-\gamma_4)}.&
\end{eqnarray}
\noindent{CLOSED CONFIGURATION}
\begin{eqnarray}\label{lc}
&\ka_{4}= \dfrac{4|G_4|^2}{\Gamma_m\Gamma_4},\quad
a_{3i}=a_{33}=1+\Delta n_4 - (1+2\Delta n_4) \dfrac{\Gamma_m-
\gamma_3}{\Gamma_m+\Gamma_n-\gamma_3},\
a_{44}=0.5[1-\dfrac{\gamma_3}{\Gamma_n}+\Delta n_3(1
+\dfrac{\gamma_3}{\Gamma_n})],&   \nonumber\\
&a_{2i}=a_{23}=1+\Delta n_3(\Gamma_n-\Gamma_m+\gamma_3)/
(\Gamma_n+\Gamma_m-\gamma_3),\
a_{14}=0.5[1+\Delta n_4(1+\gamma_3/\Gamma_n)].&
\end{eqnarray}
\centerline{\bf H-scheme (fields $ E _ 4 $, $ E _ 2 $; $ i = 2 $)}
\noindent {OPEN CONFIGURATION}
{\small\begin{eqnarray}\label{ho}
\ka_2=\dfrac{2|G_2|^2(\Gamma_f+\Gamma_m-\gamma_2)}
{\Gamma_f\Gamma_m\Gamma_2}; \
a_{14}=1;\quad  a_{2i}=a_{22}=1;\
a_{3i}=a_{32}=\dfrac{\Gamma_l-\gamma_4}{\Gamma_l} \dfrac{\Gamma_f-\gamma_2}{\Gamma_m+\Gamma_f-
\gamma_2} ;\ a_{44}=\dfrac{\Gamma_l}{\Gamma_l+\Gamma_m-\gamma_4}.
\end{eqnarray}}
\noindent{CLOSED CONFIGURATION}
\begin{eqnarray}\label{hc}
&\ka_4={4|G_4|^2}/{\Gamma_m\Gamma_4} ,\
a_{3i}=a_{32}=(1+2\Delta n_4)(\Gamma_f-\gamma_2)/(\Gamma_m+\Gamma_f-
\gamma_2) -\Delta n_4,\
a_{44}=0.5(1-\Delta n_2),&\nonumber\\
&a_{14}=0.5(1+\Delta n_4),\
a_{2i}=a_{22}=1+\Delta n_2(\Gamma_m-\Gamma_f+\gamma_2)/
(\Gamma_m+\Gamma_f-\gamma_2).&
\end{eqnarray}
$NIE$ are associated with coherence at two-photon transitions and disappear at
$|P_{4i}|\rightarrow \infty$. At $G_4=0$ formulas (\ref{alpharch}), (\ref{der})
converge into those, similar to discussed and analyzed in
\cite{{Pop},{Vved},{Izv},{Spie}}.  Following \cite{Pop}, range of parameters
where amplification is not accompanied by the inversion of power-dependent
populations can be easily found from (\ref{alpharch}), (\ref{der}). For the
resonant coupling in $V$ and $\Lambda$ schemes conditions for AWI at the
transitions 4 and $i$ take the form, correspondingly:
\begin {equation} \label{rawi}
{\Delta r_4}/{\Delta r_i}<{g_2}/{(1+u_2)}; \quad {\Delta r_i}/{\Delta
r_4}<{u_1}/{(1+g_1)}.
\end{equation}
Similar formulas for $ H $  schemes show, that on the contrary to the previous
configurations, inversion of populations on the adjacent transition is required
for $AWI$ in  center of the resonance, or amplification under certain
conditions arises in wings of a resonance. (More detail formulas and analysis
are given in \cite{Kuch}.

Threshold and output power of lasing without inversion can be found from the
equation:
\begin {equation} \label {las}
\alpha_4=T,
\end{equation}
Where $ T $ is loss of a radiation from a laser cavity per one pass, scaled
to the unit of length of the amplifying medium. Thus, the derived expressions
determine conditions and characteristics of inversionless generation too.
\subsection {\small Numerical analysis}
\begin{figure}[!h]
\epsfxsize=.9\textwidth \center{\leavevmode\epsfbox{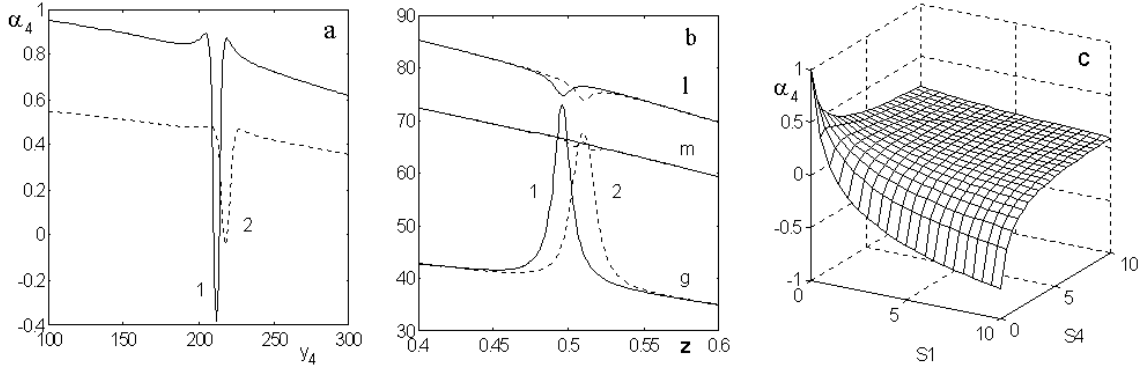}}
\vspace{-1mm}
\caption {\label{Ne}\small Velocity-averaged  absorption/gain index and populations vs
velocities at inhomogeneously broadened $l-m$ transition ($l$ -- excited state,
open configuration) in the presence of strong field at  $l-g$ transition ($Ne$,
$2s_2$ -- $2p_4$ -- $3s_2$) (co-propagating waves). $\lambda_1=1150nm$,
$\lambda_4=630nm$ ($Ne$), $ \Gamma _ m = 3\cdot 10 ^ 7 c ^ {-1} $, $ \Gamma _ l
= 5\cdot 10 ^ 7 c ^ {-1} $, $ \Gamma _ g = 10 ^ 7 c ^ {-1} $, $ \gamma _ {ml} =
\gamma _ {ml} = 0.5\cdot 10 ^ 7 c ^ {-1} $.  Ratio of initial populations is:
$N_l:N_g:N_m = 100:50:85$. {\bf a} -- absorption index
($y_4=\Omega_4/\Gamma_{lm}$), {\bf b} --  populations vs velocities ($z=v/\bar
v$, frequency $\omega_4$ is locked to the absorption minimum).  1 --
$S_1=|E_1d_{lg}/2\hbar|^2/\Gamma_{1}\Gamma_{gm}=5$,\/
$S_4=|E_4d_{ml}/2\hbar|^2/\Gamma_{4}\Gamma_{gm}=0$.
$\Omega_1=350\Gamma_{lg}$;\/ 2 -- $S_1=5$, $S_4=1$, $\Omega_1=360\Gamma_{lg}$.
{\bf c} --  velocity-averaged  absorption/gain index  vs field intensities
($\Omega_1=\Omega_4=0$).}
\end{figure}
Below we shall apply the derived expressions for numerical analysis of $NIE$
in open ($Ne$)  and closed ($Na$)  configurations of transitions.  The same
transitions of $Ne$  were considered in \cite{Pop} for illustration of
possible AWI of weak probe field. The formulas for velocity averaged
absorption index were derived and  difference in $NIE$ for backward and
forward waves in inhmogeneously broadened transitions  was analyzed in \cite
{{Sok},{Vved}} for the cases of weak probe field and coupling  Rabi frequency
of driving field not exceeding Doppler width of the  transition. Therefore in
further main attention will be given to effects accompanying increase of
intensity of an amplifying radiation.

\begin{figure}[!h]
\epsfxsize=.47\textwidth \leavevmode\epsfbox{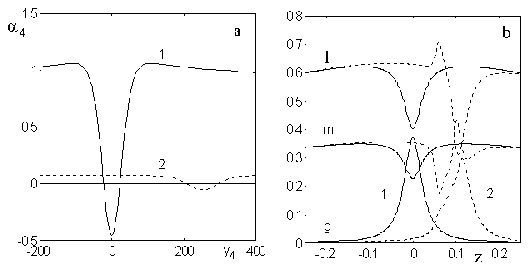}\hfill
\epsfxsize=.47\textwidth \leavevmode\epsfbox{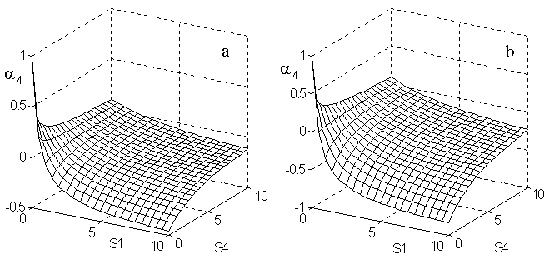}\\
\parbox{.47\textwidth}{
\caption{\label{Na}\small Velocity-averaged  absorption - gain index ({\bf a}) and populations
vs velocities ($ z = v/\bar v $) ({\bf b}) at inhomogeneously broadened $l-m$
transition  in the presence of strong field at  $l-g$ transition  ($ l $ -
ground state, closed configuration, co-propagating waves).  $ \lambda _ 4 =
330nm $, $ \lambda _ 1 = 590nm $ ($Na$; $3P _ {1/2}$ -- $ 3S $ -- $4P_{1/2}$).  $
\Gamma _ m = 9\cdot 10 ^ 7 c ^ {-1} $, $ \Gamma _ g = 63\cdot 10 ^ 7 c ^ {-1}$,
$ N _ l:N _ g:N _ m = 64:0:36 $. {\bf a}: $ 1$ -- $ S_1 = 10, S_ 4 = 0, $ $ y _
1= 0$; \, $2$ -- $S_1 = 10,  S_4 = 20 $, $ y _ 1 = 20 $.  {\bf b } --
$\omega_4$ corresponds to: $1$ -- to the right, $2$ -- to the left point of a
zero absorption of the appropriate curve in {\bf a }.}}\hfill
\parbox{.47\textwidth}{
\caption{\label{Nas}\small Velocity-averaged  absorption/gain index  vs field intensities
for co-propagating waves ($\Omega_1=\Omega_4=0$). {\bf a } --
$N_l:N_g:N_m=64:0:36$; {\bf b } --  $N_l:N_g:N_m=60:0:40$. Other parameters
are like in FIG. \protect\ref{Na}.}}
\end{figure}

FIG. \ref{Ne}{\bf a} shows that inhomogeneous broadening does not destroy macroscopic
coherence effects and $NIE$. The relative change of absorption index by an
auxiliary field appears even larger, than for a homogeneously broadened
transition.  Coherent coupling gives rise to amplification at $\omega_4$ and
establishes populations so that there is no inversion neither on one- nor on
two-photon transitions FIG. \ref{Ne}{\bf b}.  With the growth of $E_4$ populations
of $m$ and $g$ levels aim to equal magnitude, which indicates appearance of
coherent population trapping. In the latter case  a small modulation of
velocity distribution on the level $ m $ appears. Like in \cite{{Pop},{Vved}},
analysis shows that in order to attain $AWI$ certain ratio between initial (in
the absence of the  radiations) populations must be fulfilled. FIG. \ref{Ne}{\bf c}
shows that absorption (gain) strongly depends on the intensities of both
driving and probe fields.
As it was outlined, the open and closed systems differ both in possible
magnitudes of relaxation parameters and in dependence of incoherent excitation
on an intensity of the driving fields.  FIG. \ref{Na} considers the case, when
36\% of a ground state atoms are initially  excited by an incoherent pump  to a
level $ m $, that still may correspond to strong absorption at the transition
$ml$. By that strong driving field $E_1$ may produce $AWI$ for co-propagating
shorter-wavelength weak radiation at $\omega_4$, which makes approximately 50\%
from an initial absorption (FIG. \ref{Na}{\bf a}). The amplification happens in
absence of saturated population inversion  for all transitions
(FIG. \ref{Na}{\bf b}).  It is essential that  a population of a top level $ m $
depends on the strength of $E_1$ even  at zero intensities of a probe radiation
$E_4$.

Curve 2 (FIG. \ref{Na}{\bf a}) shows, that the AWI strongly depend on
 intensity of an amplified radiation, that is accompanied
for the given configuration  by noticeable change of populations
of levels $ m $ and $ l $. (FIG. \ref{Na}{\bf b}) displays energy and velocity
distribution of the atoms corresponding to appearance of transparency at
$\omega_4$. It is interesting to note that  the distribution sharply varies
with increase of intensity of $E_1$.

FIG. \ref{Nas}  shows, that $AWI$ may be maintained in a certain level with
growth of intensity of an amplified radiation by changing incoherent excitation
rate and strength of an auxiliary radiation.
\section{\small INTERFERENCE EFFECTS IN RESONANT $FWM$ AT DOPPLER BROADENED
TRANSITIONS}
The use of resonant $FWM$ in gases for frequency conversion allows one to decrease
the required power of fundamental radiations down to the magnitudes characteristic
for cw lasers \cite {{Hin},{Im},{Kl},{Bol},{Ar}}.  $FWM$ concerns to so-called
coherent nonlinear - optical processes, depending on phase-mismatch.
As it was outline above, at resonant
coupling, various coherent component stipulated by correlated quantum
transitions and giving contribution to the process of radiation conversion can
interfere.  Constructive interference gives rise to enhancement of appropriate
components of nonlinear polarization and destructive on the contrary - to
elimination. Studies of  appearances of quantum interference at $FWM$ continue
to attract significant interest \cite {Cop} in the context of possible use for
increase of conversion efficiency. The values of interfering components depend
on  energy level populations, which, in turn, depend on intensity of radiations
 and on processes of incoherent excitation and relaxation. Constructive or
destructive character of an interference depends on a relation of phases
coherent component and, therefore, on detunings from resonance and on type of
nonlinear - optical process.  In gaseous media inhomogeneous Doppler broadening
of transitions is characteristic of typical experimental conditions. Depending
on energy level, value and sign of Doppler shift,  contributions of atoms,
moving at various speeds, to macroscopic nonlinear polarization can both
enhance  and  suppress each other. The above listed effects appear in a
different way in  absorption, refraction and in $FWM$  macroscopic
polarization.  It turns out that even small incoherent excitation of the levels
at Doppler broadened transitions may drastically change spectral properties and
by orders of magnitude value of the nonlinear susceptibility
\cite{{Bu},{Vved},{WLP}}. The choice of optimal conditions of conversion is
essentially determined not only by influence of interference processes on
nonlinear susceptibilities, but also on indices of an absorption
(amplification) and refraction for coupled waves propagating through a
resonance medium.  Velocity selective population transfer and other effects of
resonant coupling with strong fields give rise to specific power saturation
effects in $FWM$.  In \cite {Hin}  experimental features  were observed, which
did not find explanations in framework of before  published lowest order
perturbation theory.
\begin{floatingfigure}{50mm}
\epsfxsize=45mm \center{\leavevmode\epsfbox{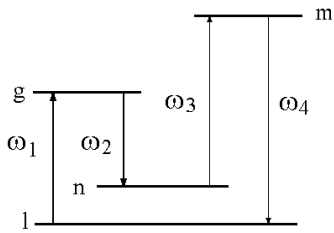}}
\caption{\label{shfwm}Coupled transitions.}
\end{floatingfigure}

This section is devoted to investigation of mutual influence of quantum
interference, relaxation, incoherent pump of levels, Doppler broadening,
effects of strong  fields on $FWM$ nonlinear polarization, and absorption of
coupled radiations in the context of experiments \cite {Hin}.
We shall consider Raman like coupling  (FIG.\ref{shfwm}) and $FWM$ processes of
a type $ \omega _ {4} = \omega _ {1} - \omega _ {2} + \omega _ {3} $  as well
as  inverse process $ \omega _ {3} = \omega _ {4} -\omega _ {1} + \omega _ {2}
$. That allows to compare influence of the above discussed elementary effects
on conversion processes in different conditions.  The elementary quantum
mechanical processes, determining $FWM$, essentially depend on that whether the
atom at an energy level couples with two strong fields or with one strong and
one weak field.

In the range of negligibly small change of the strong radiations both due to
absorption and conversion and assuming exact phase-matching, quantum efficiency
of conversion $E_3$ in $E_{4}$ $\eta_{q4}(\omega _{2}) =
(k_3/k_4)|E_4(z)/E_3|^2$  can be presented in the form:
\begin{eqnarray}\label{q4}
\eta_{q4}=k_3k_4{\mid 2\pi N\chi ^{(3)}_{4}E_1E_{2}^*
/(\Delta\alpha/2)\mid ^{2}}
exp\{-\alpha_4z\}(exp\{-(\Delta\alpha/2)z\}-1)^2,
\end{eqnarray}
where  $N$ -- atomic number density, $\chi^{(3)}_{4}$ -- nonlinear
susceptibility, $\Delta\alpha=\alpha_1+\alpha_2+\alpha_3-\alpha_4 \approx
\alpha_3-\alpha_4 $, $\alpha_j$ -- absorption index for corresponding
radiation. Quantum efficiency for conversion of $E_4$ in $E_{3}$ is written in
symmetrical form by substitution of $\chi_4^{(3)}, \alpha_4$  for
$\chi_3^{(3)}, \alpha_3$. For small length
$z<<min\{(\alpha_{4,3})^{-1}, (\Delta\alpha/2)^{-1}\}$
spectral features of conversion is determined only by nonlinear susceptibility
and by intensities of the strong fields.
\begin{equation}\label{q43}
\eta_{q4,3}=k_3k_4\mid 2\pi N\chi ^{(3)}_{4,3}E_1E_{2}\mid ^{2}z^2
\end{equation}

\subsection { $FWM$ in  two strong and one weak
fields at the conditions of maximum coherence and coherent population trapping}\label{ss1}
In this section expressions for nonlinear polarization at frequencies $ \omega
_ 3 $ and $ \omega _ 4 $ will be presented for cases, when the fields $ E _ 1 $
and $ E _ 2 $ are strong. Their frequencies $ \omega _ 1 $ and $ \omega _ 2 $
are close accordingly to the transition frequencies $\omega _ {lg} $ and
$\omega _ {ng}$.  Radiations $ E _ 3 $ and $ E _ 4 $ with frequencies $ \omega
_ 3 $ and $ \omega _ 4 $, close to the  transition frequencies  $\omega _ {nm}
$ and $\omega _ {lm} $ - are supposed nonperturbatively weak. $NIE$ in two
strong fields may give rise to such population transfer between levels $l $ and
$n$ that population of the intermediate level $g$ change negligibly.
Such behavior is similar to $CPT$  (for  review see \cite {Agap}).
In more general sense a term $CPT$ is often applied to the processes
whereas contribution of $NIE$ in population transfer is crucial.
Such type of coupling will be considered in this section.

Nonlinear $FWM$ susceptibilities $\tilde\chi ^{(3)}_{3,4}$ are determined by the
equations:
\begin{equation}
P^{NL}(\omega _{3})=N \tilde\chi ^{(3)}_{3}E_{1}^*E_2E_{4},\quad
P^{NL}(\omega _{4})=N \tilde\chi ^{(3)}_{4}E_{1}E_2^*E_{3}.
\end{equation}
Expressions for  $\tilde\chi_{3,4}$
as well as for absorption (refractive) susceptibilities $\chi_i$, derived
with density matrix in similar way as that in Section II, are  given by
\cite{MysZ}:
\begin{eqnarray}\label{r3}
&\tilde\chi_3=\dfrac{-iK}{d_3(1+q_2)}\left[
\dfrac{R_1^*}{P_1^*}\left(\dfrac1{P_{41}}+\dfrac1{P_{12}^*}\right)+
\dfrac{R_2}{P_2P_{12}^*}+\dfrac{R_4}{P_4P_{41}}\right],&\nonumber\\
&\tilde\chi_4=\dfrac{-iK}{d_4(1+q_1)}\left[
\dfrac{R_1}{P_1P_{12}}+\dfrac{R_2^*}{P_2^*}\left(\dfrac1{P_{32}}+
\dfrac1{P_{12}}\right)+\dfrac{R_3}{P_3P_{32}}\right],& \\
&\chi_i/\chi_i^0=\Gamma_i{R_i}/{P_i}{\Delta n_i}, R_1=[{(1+g_2^*)\Delta
r_1-u_1^*\Delta r_2}]/({1+g_2^*+u_2}), R_2=[{(1+u_2^*)\Delta r_2-g_3\Delta
r_1}]/({1+g_2+u_2^*}),&\nonumber\\
&R_3=\dfrac{\Delta r_3(1+q_1)-u_3R_2^*(1-q_3)+q_1u_2R_1}{1+q_1+u_4},\quad R_4=\dfrac{\Delta
r_4(1+q_2)-g_1R_1^*(1-q_4)+q_2g_2R_2}{1+q_2+g_4}.&\nonumber\\
&g_1=\dfrac{|G_1|^2}{P_{41}P_1^*}, g_2=\dfrac{|G_1|^2}{P_{12}^*P_2},
g_3=\dfrac{|G_1|^2}{P_{12}^*P_1^*}, g_4=\dfrac{|G_1|^2}{P_{41}P_4},
u_1=\dfrac{|G_2|^2}{P_{12}^*P_2}, u_2=\dfrac{|G_2|^2}{P_{12}P_1},
u_3=\dfrac{|G_2|^2}{P_{32}P_2^*}, u_4=\dfrac{|G_2|^2}{P_{32}P_3},&\nonumber\\
&q_1=\dfrac{|G_1|^2}{P_{32}d_4}, q_2=\dfrac{|G_2|^2}{P_{41}d_3},
q_3=\dfrac{|G_1|^2}{P_{12}d_4}, q_4=\dfrac{|G_2|^2}{P_{12}^*d_3}, \Delta
r_1=\dfrac{\Delta n_1X_2-\Delta n_2X_3} {X_1X_2-X_3X_4}, \Delta
r_2=\dfrac{\Delta n_2X_1-\Delta n_1X_4} {X_1X_2-X_3X_4},&\nonumber\\
& d_3=\Gamma_3+i(\Omega_4-\Omega_1+\Omega_2),
d_4=\Gamma_4+i(\Omega_1-\Omega_2+\Omega_3),  &\nonumber\\
&r_j=n_j+\Delta r_2(b_1^j\ka_1F_2+b_2^j\ka_2F_3)- \Delta
r_1(b_1^j\ka_1F_1+b_2^j\ka_2F_4),&\nonumber\\
 &X_1=1+a_1 \ka_1 F_1-a_2 \ka_2 F_4, \quad X_2=1+a_3 \ka_2 F_3-a_4\ka_1 F_2,&\nonumber\\
& X_3=a_2 \ka_2 F_3-a_1 \ka_1 F_2,\quad
X_4=a_4 \ka_1 F_1-a_3\ka_2 F_4.&\\ \label{sr}
&F_1=Re\{\dfrac{\Gamma_1}{P_1}\dfrac{1+g_2^*}{1+g_2^*+u_2}\},
F_2=Re\{\dfrac{\Gamma_1}{P_1}\dfrac{u_1^*}{1+g_2^*+u_2}\},
F_3=Re\{\dfrac{\Gamma_2}{P_2^*}\dfrac{1+u_2}{1+g_2^*+u_2}\},
F_4=Re\{\dfrac{\Gamma_2}{P_2^*}\dfrac{g_3^*}{1+g_2^*+u_2}\}.&\nonumber
\end{eqnarray}
Here and further $\Delta r_1=r_l-r_g, \Delta
r_2=r_n-r_g, \Delta r_3=r_n-r_m, \Delta r_4=r_l-r_m, \Delta n_4=n_l-n_m$,
etc., $ r_ j $ and $ n _ j $  are population  of levels accordingly power
dependent and in absence of the fields; $ \ka _ 1$, $ \ka _ 2 $ --  saturation
parameters; $ a _ k $, $ b _ k ^ j $ -
coefficients, determined by relaxation properties of the transitions,
which are different for open and closed transition configurations and
will be defined below; $K$ is constant.\\
\noindent { OPEN CONFIGURATIONS} \\
For the open system  the parameters in (\ref{sr})are:
\begin{eqnarray*}
&\ka_1=\dfrac{2|G_1|^2(\Gamma_g+\Gamma_l-\gamma_1)}
{\Gamma_g\Gamma_l\Gamma_1},\quad
\ka_2=\dfrac{2|G_2|^2(\Gamma_g+\Gamma_n-\gamma_2)}
{\Gamma_g\Gamma_n\Gamma_2},&\nonumber\\
&a_2=\dfrac{\Gamma_n}{\Gamma_l}
\dfrac{\Gamma_l-\gamma_1}{\Gamma_g+\Gamma_n-\gamma_2},\quad
a_4=\dfrac{\Gamma_l}{\Gamma_n}\dfrac{\Gamma_n-\gamma_2}
{\Gamma_l+\Gamma_g-\gamma_1},\quad
a_1= a_3=1,&\\
&b_1^g=-\dfrac{\Gamma_l}{\Gamma_l+\Gamma_g-\gamma_1},\quad
b_2^g=\dfrac{\Gamma_n}{\Gamma_g+\Gamma_n-\gamma_2} ,\quad
b_1^n=-\dfrac{\gamma_2}{\Gamma_n}b_1^g,\quad&\nonumber\\
&b_2^n=-\dfrac{\Gamma_g-\gamma_2}{\Gamma_g+\Gamma_n-\gamma_2},\quad
b_1^l=\dfrac{\Gamma_g-\gamma_1}{\Gamma_l+\Gamma_g-\gamma_1},\quad
b_2^l=\dfrac{\gamma_1}{\Gamma_l}b_2^g,\quad
b_i^m=0.&
\end{eqnarray*}
\noindent { CLOSED CONFIGURATION} \\
The corresponding parameters in  (\ref{sr})  take the values:
\begin{eqnarray*}
&\ka_1={4|G_1|^2}/{\Gamma_g\Gamma_1}, a_1=0.5[1+\Delta n_1(1+{\gamma_2}/{\Gamma_n})],
a_2=1+\Delta n_1-(1+2\Delta n_1)
({\Gamma_g-\gamma_2})/({\Gamma_g+\Gamma_n-\gamma_2}), &\\
&a_3=1+\Delta n_2[1-2({\Gamma_g-\gamma_2})/({\Gamma_g+\Gamma_n-\gamma_2})],
a_4=0.5[1-({\gamma_2}/{\Gamma_n})+\Delta n_2[1+({\gamma_2}/{\Gamma_n})], &\\
&b_1^l/n_l=b_1^m/n_m=({\Gamma_n+\gamma_2})/{2\Gamma_n},
b_2^l/n_l=b_2^l/n_l=-({\Gamma_n-\Gamma_g+\gamma_2})/
({\Gamma_g+\Gamma_n-\gamma_2}),&\\
 &  b_1^n=0.5n_n-(1-n_n){\gamma_2}/{2\Gamma_n},
b_2^n=(2n_n-1)({\Gamma_g-\gamma_2})/({\Gamma_g+\Gamma_n-\gamma_2})-n_n,&\\
&
b_1^g=n_g({\Gamma_n+\gamma_2})/{2\Gamma_n}-0.5,
b_2^g=1-n_g-(1-2n_g)({\Gamma_g-\gamma_2})/({\Gamma_g+\Gamma_n-\gamma_2}),&\\
&
n_l={(1+w_m/\Gamma_m+w_g/\Gamma_g+w'_n/\Gamma_n)}^{-1},
n_m=w_mn_l/\Gamma_m, n_g=w_gn_l/\Gamma_g, &\\
&
n_n=w_n'n_l/\Gamma_n, w_n'=w_n+w_g\gamma_{gn}/\Gamma_g+w_m\gamma_{mn}/
\Gamma_m&
\end{eqnarray*}
\subsection {$FWM$ of two strong and two weak radiations under condition of
perturbation of each resonant energy level only by one strong radiation}
\label{sw}
In \cite {Hin} the features coming out from  increase of intensity of a
radiation at frequency $ \omega _ 3 $ were investigated too. Consider a case,
where the fields $ E _ 1 $ and $ E _ 3 $ are strong, and $ E _ 2 $ and $ E _4$
- weak.  With the aid of solution of a set of equations for off- and diagonal
elements of density  matrix  up to the first order of perturbation theory in respect
of the weak fields equations for the susceptibilities can be presented as
\cite{MysK}:
\begin{eqnarray}
&\tilde\chi_2=\dfrac{-iK}{d_2(1+v_5^*+g_5^*)}
\left[\left(\dfrac{\Delta r_1}{P_1P_{41}^*}+
\dfrac{\Delta r_3}{P_3P_{43}^*}\right)+
\dfrac {R_4^*}{P_4^*}\left(\dfrac 1{P_{41}^*}+\dfrac 1{P_{43}^*}\right)\right],
v_5=\dfrac{|G_3|^2}{P_{41}d_2^*}, g_5=\dfrac{|G_1|^2}{P_{43}d_2^*},
v_1=\dfrac{|G_3|^2}{P_{43}P_3^*},
&\nonumber\\
&
\tilde\chi_4=\dfrac{-iK}{d_4(1+v_7^*+g_7^*)}
\left[\left(\dfrac{\Delta r_1}{P_1P_{12}}+
\dfrac{\Delta r_3}{P_3P_{32}}\right)+
\dfrac {R_2^*}{P_2^*}\left(\dfrac 1{P_{12}}+\dfrac 1{P_{32}}\right)\right],
v_7=\dfrac{|G_3|^2}{P_{12}^*d_4^*}, g_7=\dfrac{|G_1|^2}{P_{32}^*d_4^*},
v_2=\dfrac{|G_3|^2}{P_{32}^*P_2},
&\label{ch} \\
&R_2=\dfrac {
\Delta r_2(1+g_7+v_7)-v_3(1+v_7-g_8)\Delta r_3-g_3(1+g_7-v_8)\Delta r_1}
{(1+g_2+v_2)+[g_7+g_2(g_7-v_8)+v_7+v_2(v_7-g_8)]},
 v_8=\dfrac{|G_3|^2}{P_{32}^*d_4^*}, g_8=\dfrac{|G_1|^2}{P_{12}^*d_4^*},
v_3=\dfrac{|G_3|^2}{P_{32}^*P_3^*},
&\nonumber\\
&
R_4=\dfrac {
\Delta r_4(1+v_5+g_5)-g_1(1+g_5-v_6)\Delta r_1-v_1(1+v_5-g_6)\Delta r_3}
{(1+g_4+v_4)+[v_5+v_4(v_5-g_6)+g_5+g_4(g_5-v_6)]},
v_6=\dfrac{|G_3|^2}{P_{43}d_2^*}, g_6=\dfrac{|G_1|^2}{P_{41}d_2^*},
v_4=\dfrac{|G_3|^2}{P_{43}P_4},&\nonumber\\
&d_2=\Gamma_{ng}+i(\Omega_1+\Omega_3-\Omega_4),
\chi_{i}/\chi_{i}^0=\Gamma_{i}{\Delta r_{i}}/{P_i}{\Delta n_i}, (i=1,3),
\chi_{i}/\chi_{i}^0=\Gamma_{i}{R_{i}}/{P_i}{\Delta n_i}, (i=2,4).&
\end{eqnarray}
The rest notations are the same as in Subsection III.A.
Expressions for the populations are:\\
\noindent {OPEN CONFIGURATION}
\begin{eqnarray*}\label{eq5}
&\Delta r_1=[{(1+\ka_3)\Delta n_1+b_1\ka_3 \Delta n_3}]/
[{(1+\ka_1)(1+\ka_3)-a_1\ka_1b_1\ka_3}],
&\\
&\Delta r_3=[{(1+\ka_1)\Delta n_3+a_1\ka_1 \Delta n_1}]/
[{(1+\ka_1)(1+\ka_3)-a_1\ka_1b_1\ka_3}],&\\
& \Delta r_2=\Delta n_2-b_2\ka_3\Delta r_3-a_2\ka_1\Delta r_1,\
\Delta r_4=\Delta n_4-a_3\ka_1\Delta r_1-b_3\ka_3\Delta r_3,&\\
&r_m=n_m+(1-b_2)\ka_3\Delta r_3,
r_g=n_g +(1-a_3)\ka_1\Delta r_1,&\\
&r_n=n_n-b_2\ka_3\Delta r_3+a_1\ka_1\Delta r_1,
r_l=n_l-b_1\ka_3\Delta r_3+a_3\ka_1\Delta r_1,&\\
&\ka_1=\ka_1^0\dfrac{\Gamma_{lg}^2}{|P_1|^2},\
\ka_1^0=\dfrac{2(\Gamma_l+\Gamma_g-\gamma_{gl})}
{\Gamma_l\Gamma_g\Gamma_{lg}}|G_1|^2,\
\ka_3=\ka_3^0\dfrac{\Gamma_{mn}^2}{|P_3|^2},\
\ka_3^0=\dfrac{2(\Gamma_m+\Gamma_n-\gamma_{mn})}
{\Gamma_m\Gamma_n\Gamma_{mn}}|G_3|^2, & \\
&a_1=\dfrac{\gamma_{gn}a_2}{\Gamma_n-\gamma_{gn}}=
\dfrac{\gamma_{gn}\Gamma_la_3}{\Gamma_n(\Gamma_g-\gamma_{gl})}=
\dfrac{\gamma_{gn}\Gamma_l}{\Gamma_n(\Gamma_l+\Gamma_g-\gamma_{gl})},&\\
&b_1=\dfrac{\gamma_{ml}\Gamma_nb_2}{\Gamma_l(\Gamma_m-\gamma_{mn})}
=\dfrac{\gamma_{ml}b_3}{\Gamma_l(\Gamma_l-\gamma_{ml})}
=\dfrac{\gamma_{ml}\Gamma_n}{\Gamma_l(\Gamma_m+\Gamma_n-\gamma_{mn})}.
&
\end{eqnarray*}
\noindent {CLOSED CONFIGURATION} \\
The populations of levels are described by the equations:
\begin{eqnarray*}\label{syst}
&\Gamma_mr_m=w_mr_l-2\Re\left\{iG_3^*r_3 \right\},\quad
\Gamma_gr_g=w_gr_l-2\Re\left\{iG_1^*r_1 \right\},\nonumber &\\
&\Gamma_nr_n=w_nr_l+2\Re\left\{iG_3^*r_3 \right\}
+\gamma_{gn}r_g+\gamma_{mn}r_m,\quad r_l=1-r_m-r_g-r_n,\nonumber &
\end{eqnarray*}
where  $r_1=iG_1\Delta r_1/P_1$, $r_3=iG_3\Delta r_3/P_3$. The solution is
\begin{eqnarray*}\label{solve}
&r_l=n_l(1+\ka_3)(1+\ka_1)/\beta,\quad
r_g=(1+\ka_3)[n_l(1+\ka_1)-\Delta n_1]/\beta,\nonumber &\\
&r_n=\left\{n_m(1+\ka_3)(1+\ka_1)+[\Delta n_3(1+\ka_1)+\Delta
n_1\gamma_2\ka_1/\Gamma_n]
(1+b\ka_3)\right\}/\beta\nonumber &\\
&r_m=\left\{n_m(1+\ka_3)(1+\ka_1)+[\Delta n_3(1+\ka_1)+\Delta
n_1\gamma_2\ka_1/\Gamma_n] b\ka_3\right\}/\beta,\nonumber &\\
&\Delta r_1=r_l-r_g=\Delta n_1(1+\ka_3)/\beta,\qquad \Delta
r_3=r_n-r_m=[\Delta n_3(1+\ka_1)+\Delta
n_1\gamma_2\ka_1/\Gamma_n]/\beta,\nonumber &
\end{eqnarray*}
where $\beta=(1+\ka_3)[1-\Delta
n_3+2(n_l+n_m)\ka_1]+(1+2b\ka_3)[\Delta n_3(1+\ka_1)+\Delta
n_1\gamma_2\ka_1/\Gamma_n]$, $\Delta n_1=n_l-n_g$, $\Delta
n_3=n_n-n_m$, $n_m=n_lw_m/\Gamma_m$, $n_g=n_lw_g/\Gamma_g$,
$n_n=n_l{w_n}'/\Gamma_n$, $n_l=(1+
w_m/\Gamma_m+w_g/\Gamma_g+{w_n}'/\Gamma_n)^{-1}$,
${w_n}'=w_n+w_g\gamma_{gn}/\Gamma_n+w_m\gamma_{mn}/\Gamma_n$,
$b=\Gamma_n/(\Gamma_m+\Gamma_n-\gamma_3)$,
$\ka_1=({2|G_1|^2}/{\Gamma_1\Gamma_g})({\Gamma_1^2}/{|P_1|^2})$,
$\ka_3=({2|G_3|^2(\Gamma_m+\Gamma_n-\gamma_3)}/{\Gamma_m\Gamma_n\Gamma_3})(
{\Gamma_3^2}/{|P_3|^2})$.
The remaining denotations are former.
\subsection {\label {dop}\small Effect of Doppler broadening on resonant $FWM$}
Formula for $\tilde\chi_4^{(3)}$  in lowest order of
perturbation theory \cite{Bu} can be derived from (\ref{r3}), (\ref{ch}) at
$G_i\rightarrow 0$:
\begin{eqnarray}
&\tilde\chi_4^{(3)}(\omega _{4}=\omega _{1}-\omega _{2}+\omega _{3})=
\dfrac{iK}{ \Gamma _{ml}+i(\Omega ^{'}_{1}- \Omega ^{'}_{2}
+ \Omega ^{'}_{3})}\cdot & \nonumber \\
&\cdot\left \{\dfrac1{\Gamma _{gm}+i(\Omega^{'}_{3}- \Omega^{'}_{2})}
\cdot [\dfrac{n_{g} - n_{n}}{\Gamma _{ng}-i\Omega ^{'}_{2}}+
\dfrac{n_{m} - n_{n}}{\Gamma _{mn}+i\Omega ^{'}_{3}}]+
\dfrac1{\Gamma _{ln}+i(\Omega ^{'}_{1}- \Omega ^{'}_{2})}
\cdot [\dfrac{n_{g} - n_{n}}{\Gamma _{ng}-i\Omega ^{'}_{2}}+
\dfrac{n_{g} - n_{l}}{\Gamma _{lg }+i\Omega ^{'}_{1}}]\right\},&
\label{eq11}
\end{eqnarray}
where
$\Omega ^{'}_{j}=  \Omega _{j} - {\bf k}_{j}{\bf v}$,
$n_{i}= N_i\cdot exp\{-({\bf v}/\bar v)^2\}/\sqrt\pi \bar v $.
As the function of $v$ all terms in (\ref{eq11}), besides those proportional
to $n_{g}- n_{n}$, have all poles in one and the same complex half plane.
Therefore at $\Gamma_i<<k_i\bar v$ only terms, proportional to $n_{g} - n_{n}$,
do not vanish after averaging over Maxwell's velocity distribution.
Velocity averaged susceptibility is:
\begin{eqnarray}
&<\tilde\chi_4^{(3)}(\omega _{4})>_{v} = \dfrac{iK\pi ^{1/2}\exp \{-(\Omega
_{2}/k_{2}\bar v)^{2}\}(N_{g} -N_{n})}
{k_{2}\bar v [\tilde{\Gamma }_{1}+i(\Omega _{1}- k_{1} \Omega _{2}/k_{2})]
[\tilde{\Gamma }_{3}+i(\Omega _{3}- k_{3} \Omega _{2}/k_{2})]},&\label{eq13}\\
&\tilde{\Gamma }_{1} = \Gamma _{nl}+ ({k_{1}/ k_{2}} - 1)\Gamma _{ng},
\quad \tilde{\Gamma }_{3} = \Gamma _{gm}+ ({k_{3}/ k_{2}} - 1)
\Gamma _{ng}.&\nonumber
\end{eqnarray}
As it is seen from (\ref{eq13}), for the process $ \omega _ {4} = \omega _
{1} - \omega _ {2} + \omega _ {3} $ interference of contributions of atoms at
different velocities to the velocity averaged nonperturbed $FWM$ nonlinear
susceptibility  $ <\tilde \chi _ 4> _ v $ leads to the fact, that in the lowest order
on the small parameter $ \Gamma _ 2/k _ 2\bar v $, it is proportional to the velocity integrated
difference between populations of the excited states $ N _ {g} -N _ {n}$.
In the similar way, with aid of (\ref{ch}) one can find, that on the
contrary, $ < \tilde\chi _ 2> _ v $ for the process $ \omega _ {2} = \omega _ {1} -
\omega _ {4} + \omega _ {3} $ in the same approximation  is determined by the
population difference on transitions from the lowest level. For the resonant
sum frequency $FWM$ $ \omega _ {4} = \omega _ {1} + \omega _ {2} + \omega _ {3}
$ in the cascade configuration of levels velocity averaged susceptibility
occurs proportional to higher order of the small parameter $ \Gamma /k\bar v $
compared to  Raman-type difference-frequency coupling \cite{Vved}.
These features demonstrate great difference between resonant $FWM$ processes in
homogeneously and inhomogeneously broadened transitions.

In strong electromagnetic fields above mentioned processes are accompanied
by the velocity selective population transfer and by some other intensity
dependent effects. In \cite {Hin} experiments on resonant $cw$ $FWM$ at
Raman-like electronic molecular transitions of $Na_2$ have been carried out.
Frequency tunable radiation at $ \omega _ 2 $ was generated at the same
transition of $Na_2$ either in external dimer Raman laser or $FWM$ was
performed inside the Raman laser cavity.  Frequency $ \omega _ 2 $ was tuned by
tuning $ \omega _ 1 $. Radiation at $ \omega _ 3 $ was provided from $cw$ dye
laser. High conversion efficiency have been attained in single frequency nearly
power saturation regime.  Observed $FWM$ frequency tuning characteristics occured
in disagreement with the predictions of lowest order perturbative theory.  From
that the authors derived the questions to be answered with the aid of an
 advanced nonperturbative theory. We shall use above presented expressions for
the numerical analysis of the models with the parameters, close to that in the
experiments, in order to explain main observed features.

The electronic - vibration-rotation transitions between $ X $, $ A $ and $ B $
electronic levels of the  dimer were used in the experiments, the lowest
electronic level being ground one. Two $FWM$ processes were investigated:  when
frequency $\omega _ 3 $ less  than $\omega _ 2 $ and, therefore frequency of a
generated radiation $ \omega _ 4 $ was less than $ \omega _ 1 $ (down
conversion) and opposite upconversion process.  As it was discussed above,
different appearances of interference processes at Doppler broadened
transitions can be expected in those cases. Main observed experimental
dependencies, which did not find explanations, can be summarized as follows.

According to ({\ref{eq13}), at $ \Omega _ {1} - k_ {1} \Omega _ {2}
/k _ {2} = 0 $, in the lowest order of perturbation theory the maximum output
of $FWM$ at $ \omega _ 4 $  as a function of $ \omega _ 3 $ corresponds to $
\Omega _ {3} =  k _ {3} \Omega _ {2} /k _ {2} = 0 $.  Lineshape of the
resonance is  Lorentzian with the linewidth of the order of characteristic
homogeneous widths of optical transitions. However, in the down conversion
experiments the wide resonance of the order of Doppler width of transition $ ml
$ with the center being locked at $ \omega _ 3 \approx \omega _ {mg} $ was
observed. It's position practically did not vary at tuning $ \omega _ {2} $
within Doppler resonance of the transition $gn$ (at the expense of tuning of $
\omega _ 1 $, so that $ \Omega _ {2} = k _ {2} \Omega _ {1} /k _ {1} $).  In
the upconversion experiments the resonance was tunable by tuning frequency  $
\omega _ {2} $, but with the slope less than  $ d\Omega _ {3} /d\Omega _ {2} =
k _ {3} /k _ {2} $.  Width of the resonance also was  commensurable with the
Doppler width of the transition $ ml $.

For numerical analysis we have used a model with the transitions
 parameters, close to those from the experiment. \\
1. Down conversion:  $ \lambda _ {ml} $ = 598 nm, $ \lambda _ {gl} $ = 488 nm,
$ \lambda _ {mn} $ = 655 nm, $ \lambda _ {gn} $ = 525 nm; $ k _ {1} \bar v =
6.94 $, $ k _ {2} \bar v = 6.45 $, $ k _ {3} \bar v = 5.17 $, $ k _ {4} u =
5.66 $, (in terms of $ 10 ^ {9} $ $ c ^ {-1} $); $ \Gamma _ {m} = 200 $, $
\Gamma _ {n} = 30 $, $ \Gamma _ {g} = 260 $,   $ \gamma _ {mn} = 2 $, $ \gamma
_ {ml} = 4 $, $ \gamma _ {gn} = 20 $, $ \gamma _ {gl} = 10 $, $ \Gamma _ {ln} =
40 $, $ \Gamma _ {nm} = 110 $, $ \Gamma _ {lm} = 110 $, $ \Gamma _ {gm} = 130
$, $ \Gamma _ {ng} = 140 $, $ \Gamma _ {lg} = 140 $, (in terms of $ 10 ^ {6} $
$ c ^ {-1} $) $ N _ l/N _ n = 30/2 $.

2. Up-conversion:  $ \lambda _ {ml} $ = 473 nm, $ \lambda _ {gl} $ = 661 nm, $
\lambda _ {mn} $ = 514 nm, $ \lambda _ {gn} $ = 746 nm; $ k _ {1} \bar v = 5.12
$, $ k _ {2} \bar v = 4.54 $, $ k _ {3} \bar v = 6.59 $, $ k _ {4} \bar v =
7.16 $, (in terms of $ 10 ^ {9} $ $ c ^ {-1} $); $ \Gamma _ {m} = 200 $, $
\Gamma _ {n} = 30 $, $ \Gamma _ {g} = 260 $,  $ \gamma _ {mn} = 2 $, $ \gamma _
{ml} = 4 $, $ \gamma _ {gn} = 20 $, $ \gamma _ {gl} = 10 $, $ \Gamma _ {ln} =
40 $, $ \Gamma _ {nm} = 110 $, $ \Gamma _ {lm} = 110 $, $ \Gamma _ {gm} = 130
$, $ \Gamma _ {ng} = 140 $, $ \Gamma _ {lg} = 140 $, (in terms of $ 10 ^ {6} $
$ c ^ {-1} $), $ N _ l/N _ n = 110/3.2 $, population of two upper levels being
negligibly small.

First, with an aid of these models  we shall illustrate a role of an
interference at velocity averaging of nonlinear susceptibilities in weak
fields.  For down conversion  in exact one- and multiphoton resonances and
homogeneously broadened transitions computing gives the ratio of squared
modulus of nonlinear susceptibilities $ | \chi _ 3 ^ {(3)} /\chi _ 4 ^ {(3)} |
^ 2 $ = 2.5. For averaged values it yields $ | < \chi _ 3 ^ {(3)} > _ v / <\chi
_ 4 ^ {(3)} > _ v | ^ 2 $ = $ 2.31\cdot10 ^ 2$.  If to change the  population
ratio for the inverse magnitude ($ N _ n/N _ l = 30/2 $), we obtain: $ | \chi _
3 ^ {(3)} /\chi _ 4 ^ {(3)} | ^ 2 $ = 0.4. The difference between averaged
values sharply decreases.  Their ratio in this case yields: $ | < \chi _ 3 ^
{(3)} > _ v / <\chi _ 4 ^ {(3)} > _ v | ^ 2 $ = 0.13.

For up-conversion similar computations give: $ | \chi _ 3 ^ {(3)} /\chi _ 4 ^
{(3)} | ^ 2 $ = 2.7, $ | < \chi _ 3 ^ {(3)} > _ v / <\chi _ 4 ^ {(3)} > _ v | ^
2$ = $ 1.45\cdot10 ^ 3 $.  At the inverse population ratio ($ N _ n/N _ l = 110/3.2 $)
we obtain:  $ | \chi _ 3 ^ {(3)} /\chi _4 ^ {(3)} | ^ 2 $ = 0.37. The difference
between averaged values sharply decreases. Their ratio in this case is:
$ | < \chi _ 3 ^ {(3)} > _ v / <\chi_ 4 ^ {(3)} > _ v | ^ 2 $ = 0.24.

Thus, effect of inhomogeneous broadening of the resonant transitions on $FWM$
processes may be very strongly dependent  on a specific process, as well as on
distribution of the populations over levels and velocities.  Therefore one can
expect  that velocity selective population transfer and other effects of strong
fields may change conclusions of the lowest order perturbative theory.

For small number density and medium length $FWM$ conversion efficiency of weak
radiation is proportional to a product of intensities of the strong radiations
and squared modulus of velocity averaged nonlinear susceptibility (equation
(\ref{q43})). The later is intensity dependent too.  Further we shall numerically
analyze effects of the strong fields on conversion efficiency with an aid of
the expressions of the subsections III.A and III.B.  Intensities of
the radiations will be characterized by the parameters $ S _ 1 = | G _ 1 | ^
2/\Gamma _ {gl} \Gamma _ {nl} $ and $ S _ 2 = | G _ 2 | ^ 2/\Gamma _ {gn}
\Gamma _ {nl} $, which are chosen in near saturation range like in the
experiments.
\begin{figure}[h!]
\epsfxsize=.7\textwidth \center{\leavevmode\epsfbox{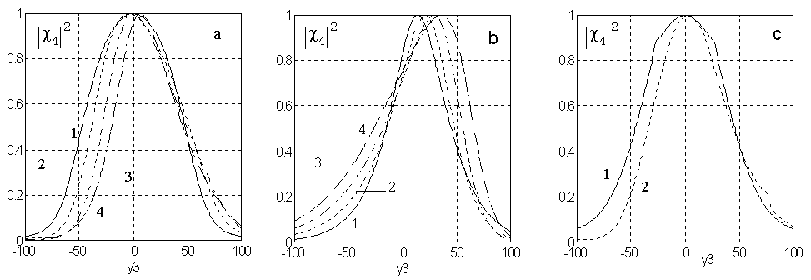}}
\caption{\label{fwmbrss}\small Squared modulus of scaled velocity averaged nonlinear
susceptibility $ | < \chi ^{(3)} _ 4> _ v/< \chi ^{(3)} _ {40}> _ v | ^ 2 $ vs scaled
detuning $ y _ 3 = \Omega _ 3/\Gamma _ {nm} $ at various detunings   $ \Omega _ 1 $ ($
\Omega _ 2 = k _ 2/k _ 1\Omega _ 1 $.  {\bf a} and {\bf b}: upconversion, $ S _ 1 =
150 $, $ S _ 2 = 350 $);  {\bf a}: 1 -- $\Omega_1=0$, 2 -- $\Omega_1=10\Gamma_{lg}$,
3 -- $\Omega_1=20\Gamma_{lg}$, 4 -- $\Omega_1=30\Gamma_{lg}$; {\bf b}: 1
-$\Omega_1=40\Gamma_{lg}$, 2 -- $\Omega_1=50\Gamma_{lg}$, 3 --
$\Omega_1=60\Gamma_{lg}$, 4 -- $\Omega_1=70\Gamma_{lg}$. {\bf c}: downconversion;  $
S _ 1 = 1000 $, $ S _ 2 = 2000 $; 1 - $ \Omega _ 1 = 0 $, 2 - $ \Omega _ 1 = 50\Gamma
_ {lg} $. }
\end{figure}
\begin{figure}[h!]
\epsfxsize=.7\textwidth \center{\leavevmode\epsfbox{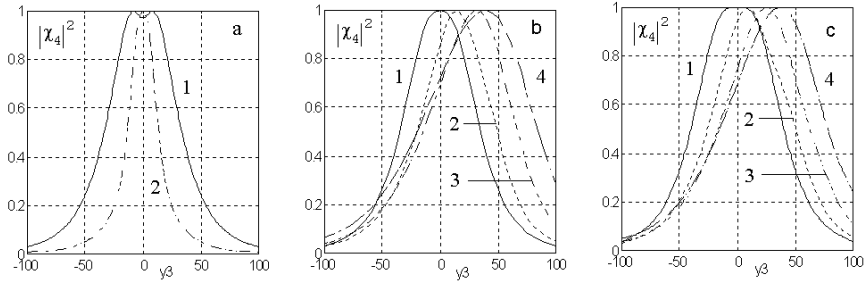}} \caption
{\label{fwmbrsw}\small Squared modulus of scaled velocity averaged nonlinear
susceptibility $ | < \chi ^{(3)} _ 4> _ v/< \chi ^{(3)} _ {40}> _ v | ^ 2 $ vs scaled
detuning $ y _ 3 = \Omega _ 3/\Gamma _ {nm} $ at various detunings   $ \Omega _ 1 $ ($
\Omega _ 2 = k _ 2/k _ 1\Omega _ 1 $) (upconversion, each level is coupled to only
one strong field).  {\bf a}: $\Omega_1=0$, $ S _ 1 = 1000 $; 1 -- $ S _ 3 = 1000 $, 2
-- $ S _ 3 = 1600 $. {\bf b}: $ S _ 1 = 2000 $, $ S _ 3 = 2100; $ 1 --  $y_1=0$, 2 --
$y_1=20$, 3 -- $y_1=40$, 4 -- $y_1=60$. {\bf c}:  same as in {\bf b}, but $ S _ 1 =
4000 $, $ S _ 3 = 4200 $. }
\end{figure}
 Figures \ref{fwmbrss} show, that for the chosen parameters due to power
broadening the resonance is much broader than homogeneous transition width and
is commensurable with Doppler linewidth, which is  of the order 60 in the used
scale.  For the parameters, corresponding to the up-conversion experiments
(FIG. \ref{fwmbrss} {\bf a,b}) in the range of small detunings $ \Omega _ 1 $
($ \Omega _ 2 = (k _ 2/k _ 1) \Omega _ 1 $) (FIG. \ref{fwmbrss} {\bf a}) the
peak of the tuning curve is displaced very insignificantly  (and even in the
opposite side, depending on the value of $ \Omega _ 1 $). At further increase of
$ \Omega_ 1 $ (FIG. \ref{fwmbrss} {\bf b}) the maximum  shifts with the
increase of $ \Omega _ 1 $, so that the slope $ \Omega _ 3/\Omega _ 1 $ is
variable. A maximum of the slope corresponds to the  detunings $ \Omega _ 1 $
of about a half of Doppler width of the transition $gl$. Thus for the
considered intensities the value of the slope reaches $\approx 0.8$, that
makes $\approx 0.5  (k _ 3/k _ 1) $.

FIG. \ref{fwmbrss} {\bf c} is computed and drown for the parameters,
corresponding to down conversion experiments.  For the considered intensities
the peak occurs locked  to the center of transition $ ml $ practically in all
an interval of  $ \Omega _ 1 $ within the Doppler width of transition $ gl $.

When  the weak field detunings are fixed and
driving fields frequencies $\omega_1 $ and $\omega_2 $ ($ \Omega _ 2 = (k _ 2/k
_ 1) \Omega _ 1 $) are tuned to the maximum, computer analysis of the slope $
d\Omega _1/d\Omega _ 3 $ in the range of $S_1=65, S_2=2.33\cdot S_1$ shows
behavior similar to that observed in experiments \cite {Hin} ($ d\Omega
_1/d\Omega _ 3 \approx k_1/2k_3$).  (Ratio  $S_1/S_2=2.33 $ is chosen according
to the ratio of prodacts of experimental field intensities and Franck-Kondon
factors.)

In the experiments \cite {Hin} the features, following increase of intensity of
$E _ 3 $, were observed too. In order to consider effects of this field and to
understand whether $CPT$ play a decisive role in observed dependencies we have
carried out numerical analysis of the up-conversion model with aid of formulas
from the Subsection III.B (FIG.  \ref{fwmbrsw}).

FIG. \ref{fwmbrsw} {\bf a} shows that at certain ratio of intensities  even
power narrowing of Doppler broadened $FWM$ resonance may happen.
FIG. \ref{fwmbrsw} {\bf b} displays approximately constant slope  $ d\Omega _
3/d\Omega _ 1 $ ($\Omega _ 3$ corresponds to the maximum output), which is
about 0.75, in a quite wide interval of $ \Omega _ 1 $.  Only for $ \Omega _ 1
$, larger than  Doppler width, the slope starts decreasing. At larger
intensities the slope varies more considerable, when $ \Omega _ 1 $ is tuned
within the Doppler line.  So for FIG. \ref{fwmbrsw} {\bf c} the slope makes up
0.1 in the vicinity of $ y _ 1 = 20 $; 0.8 - in the range of $ y _ 1 = 40 $ and
0.6 - at $ y _ 1 = 60 $.
\begin{figure}[!h]
\epsfxsize=.9\textwidth \center{\leavevmode\epsfbox{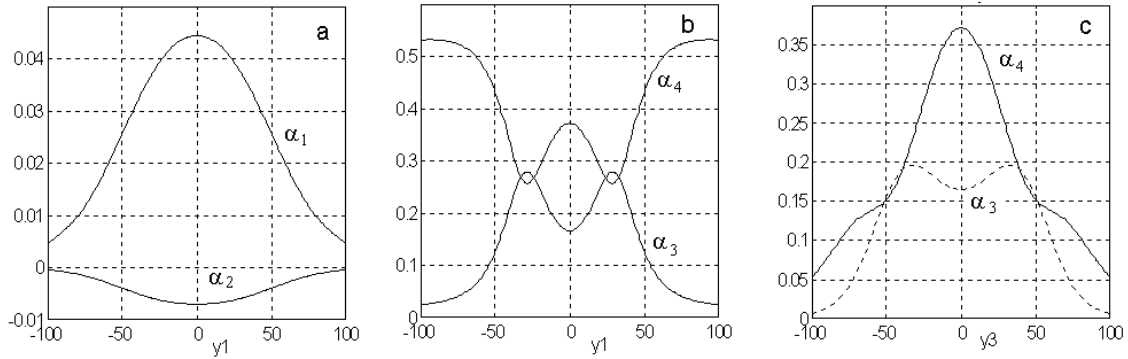}}
 \caption {\label{fwmab}\small Velocity-averaged  absorption/gain indices
scaled to the nonperturbed absorption index at $ \omega = \omega _ {gl} $
$<\alpha_i>_v/<\alpha_1^0>_v$ at $ S _ 1 = 150 $, $ S _ 2 = 350 $ (upconversion). {\bf a}  --
absorption of strong radiations vs $ \omega _ 1 $ ($ \Omega _ 2 = k _
2/k _ 1\Omega _ 1 $);  change in absorption of converting  and generated
radiations as tuning  $ \omega _ 1 $ ($ \Omega _ 3 = 0 $) ({\bf b })
and  $ \omega _ 3 $ ($ \Omega _ 1 = 0 $) ({\bf c }).}
\end{figure}

\begin{floatingfigure}{55mm}
\epsfxsize=50mm \center{\leavevmode\epsfbox{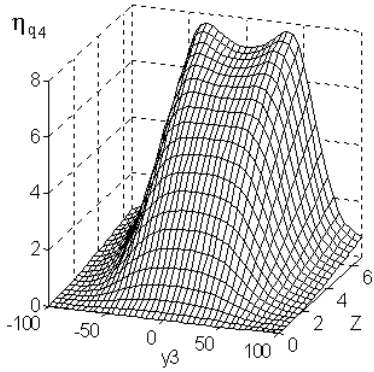}}
\caption{\label{fwmpr}\small Effect absorption on $FWM$.}
\end{floatingfigure}
In an absorbing medium spectral properties of absorption indices
may bring important effect on $FWM$.   FIG. \ref{fwmab} {\bf a-c} display
corresponding lineshapes.  FIG. \ref{fwmpr} is computed with aid of formula
(\ref{q4}) and shows that additional broadening of the tuning curve of $FWM$
output may arise from the propagation effect in an absorbing medium
($Z=\alpha_{10}z$, $\Omega_1 = 0, S_1 = 150, S_2 = 350$ (upconversion)).

In conclusion, the theory of nonlinear interference processes at Doppler
broadened quantum transitions in two strong resonant optical fields is
developed.  The derived formulas allow one account for such contributing
processes relevant to experiments as various relaxation channels, incoherent
excitation by an external source, population transfer and other coherent and
incoherent effects accompanying coupling with strong optical fields in various
open and closed configuration of transitions. Explicit formulae accounting for
the effects of the strong fields are derived.  Such appearance of quantum
interference as amplification (and lasing) without inversion of power
saturated populations and specific effects in resonant four wave mixing in
gases are analyzed with the aid of numerical models. Parameters
of the model are close to those of some recently carried out experiments. Crucial
effect of Doppler-broadening  on the contributions of the populations of
different levels to four-wave mixing, which determine selection of optimal energy-level
configuration and conditions for the experiments, is shown. Unexpected
dependencies, observed in the experiments are explained.\\

\centerline{\bf ACKNOWLEDGMENTS}
\vspace {1ex}
The author would like to thank S. A. Myslivets and
 V. M. Kuchin for assistance in calculations,  A. A. Apolonskii, S. A. Babin, U. Hinze
and B. Wellegehausen for useful discussion of their experiments prior
publication.  This work was supported by the  Russian Foundation for Basic
Research and by the Deutsche Forschungsgemeinschaft through collaborative Grant
96-02-00016 G.  The author wishes to express his sincere thanks to
L. J. F. Hermans from Hygens Laboratorium and  the Netherlands Science Foundation
(NWO) for support of this work.

\end{document}